\documentclass[aps, prl, twocolumn, superscriptaddress, showpacs]{revtex4}
\usepackage{graphicx}
\usepackage{array}
\usepackage{amsmath,amssymb}
\begin{document}
\title{Spin susceptibility of Anderson impurities in arbitrary conduction bands}
\author{Tie-Feng Fang}
\email{fangtiefeng@lzu.edu.cn}
\affiliation{Center for Interdisciplinary Studies and Key Laboratory for Magnetism and Magnetic Materials of MOE, Lanzhou University, Lanzhou 730000, China}
\author{Ning-Hua Tong}
\affiliation{Department of Physics, Renmin University of China, Beijing 100872, China}
\author{Zhan Cao}
\affiliation{Center for Interdisciplinary Studies, School of Physics, Lanzhou University, Lanzhou 730000, China}
\author{Qing-Feng Sun}
\affiliation{International Center for Quantum Materials, School of Physics, Peking University, Beijing 100871, China}
\affiliation{Collaborative Innovation Center of Quantum Matter, Beijing 100871, China}
\author{Hong-Gang Luo}
\affiliation{Center for Interdisciplinary Studies, School of Physics, Lanzhou University, Lanzhou 730000, China}
\affiliation{Beijing Computational Science Research Center, Beijing 100084, China}
\date{\today}
\begin{abstract}
Spin susceptibility of Anderson impurities is a key quantity in understanding the physics of Kondo screening. Traditional numerical renormalization group (NRG) calculation of the impurity contribution $\chi_{\textrm{imp}}$ to susceptibility, defined originally by Wilson in a flat wide band, has been generalized before to structured conduction bands. The results brought about non-Fermi-liquid and diamagnetic Kondo behaviors in $\chi_{\textrm{imp}}$, even when the bands are not gapped at the Fermi energy. Here, we use the full density-matrix (FDM) NRG to present high-quality data for the local susceptibility $\chi_{\textrm{loc}}$ and to compare them with $\chi_{\textrm{imp}}$ obtained by the traditional NRG. Our results indicate that those exotic behaviors observed in $\chi_{\textrm{imp}}$ are unphysical. Instead, the low-energy excitations of the impurity in arbitrary bands only without gap at the Fermi energy are still a Fermi liquid and paramagnetic. We also demonstrate that unlike the traditional NRG yielding $\chi_{\textrm{loc}}$ less accurate than $\chi_{\textrm{imp}}$, the FDM method allows a high-precision dynamical calculation of $\chi_{\textrm{loc}}$ at much reduced computational cost, with an accuracy at least one order higher than $\chi_{\textrm{imp}}$. Moreover, artifacts in the FDM algorithm to $\chi_{\textrm{imp}}$, and origins of the spurious non-Fermi-liquid and diamagnetic features are clarified. Our work provides an efficient high-precision algorithm to calculate the spin susceptibility of impurity for arbitrary structured bands, while negating the applicability of Wilson's definition to such cases.
\end{abstract}
\maketitle

\section{I. introduction}
The understanding of quantum impurity systems is of prime importance in condensed matter physics. Such systems, consisting of a small subsystem (the impurity) with only few degrees of freedom coupled to a continuous bath of noninteracting particles, often exhibit rather complex behavior due to the nontrivial interplay between Coulomb repulsion at the impurity site and the impurity-bath interaction. A typical example is the Kondo effect \cite{Hewson1993} describing the screening of a localized magnetic moment by conduction electrons. Kondo screening mechanisms are mostly characterized by the spin susceptibility of the impurity, a fundamental quantity describing the impurity's magnetic response. Quantum impurities, such as magnetic molecules, qubits, adatoms, or quantum dots, would also constitute an enormous miniaturization of data processing technology by encoding and storing information in their magnetic states \cite{App}. Making this prospect feasible requires very accurate knowledge on the local magnetic response of the impurity. Experimentally, such local magnetic properties are relevant in traditional nuclear magnetic resonance and neutron scattering experiments \cite{Hewson1993}.

Theoretically, a prototype model of quantum impurities is the single-impurity Anderson model \cite{Anderson1961}. While exact solutions of the Anderson model are available by the Bethe ansatz \cite{BA1983} in some limiting cases, Wilson's numerical renormalization group (NRG) \cite{NRG1980,Bulla2008} provides a systematic nonperturbative method for arbitrary impurity systems. For the impurity magnetic response in the Kondo physics, the NRG calculation of the Anderson model often emphasizes Wilson's definition \cite{NRG1980} of the impurity contribution $\chi_\textrm{imp}$ to the susceptibility due to a global magnetic field (in the following, we term it the impurity susceptibility as usual), rather than the standard definition \cite{Bulla2008} of the local spin susceptibility $\chi_\textrm{loc}$ in response to a local field. Despite the experimental relevance, this is because evaluation of the local susceptibility is equivalent to calculating a dynamical spin correlation function which is much more complex and less accurate in the traditional NRG \cite{NRG1980,Bulla2008,TNRG} or its reduced density-matrix extension \cite{Hofstetter2000}. Alternatively, one may calculate $\chi_\textrm{loc}$ by differentiating the local magnetization with respect to the external field \cite{Anders2006,Merker2012}. From a numerical point of view, however, performing differentiations is something to avoid if possible. Since in the most relevant case of a flat wide band the impurity susceptibility $\chi_\textrm{imp}$ exactly coincides with the local susceptibility $\chi_{\textrm{loc}}$, it is clearly more convenient to only calculate $\chi_\textrm{imp}$ which is a thermodynamic quantity and thus can be very accurately obtained using the traditional NRG.

However, the nature of the electron bath in the Kondo problem varies from one realization to another. It may possess a band of strong energy dependence, as in magnetic impurities adsorbed on graphene \cite{Mastrogiuseppe2014} and square lattice \cite{Zhuravlev2011}, double quantum dots \cite{Silva2006}, the narrow-band Anderson model \cite{Hofstetter1999}, and so on, for which the two susceptibilities may differ significantly \cite{Santoro1991}. In this case, extreme caution should be exercised when applying Wilson's definition, because the impurity susceptibility, being defined through global quantities, involves a subtlety from the conduction bath such that it may not necessarily yield the intrinsic spin dynamics of the impurity. Particularly, a negative $\chi_\textrm{imp}$ was found by the calculations \cite{Hofstetter1999,Silva2006,Zhuravlev2011,Mastrogiuseppe2014}, signaling a diamagnetic Kondo impurity, even with non-Fermi-liquid behavior violating the low-temperature plateau in $\chi_{\textrm{imp}}$ \cite{Mastrogiuseppe2014,Zhuravlev2011}. This is inconsistent with one's intuition. We emphasize that in those systems \cite{Hofstetter1999,Silva2006,Zhuravlev2011,Mastrogiuseppe2014}, though strongly energy dependent, the conduction bands are not gapped at the Fermi energy. Therefore, the impurity Kondo physics is very likely to be still conventional, without the exotic non-Fermi-liquid and diamagnetic behaviors (gapped bands \cite{Fang2013} could, of course, lead to non-Fermi-liquid behavior). With this respect, in order to provide a definitive answer to whether an energy-dependent band without gap could indeed make the impurity diamagnetic and a non-Fermi liquid, it is highly desirable to establish a dynamical method of precisely calculating the local susceptibility. This is because $\chi_{\textrm{loc}}$ is the standard quantity \cite{Bulla2008} that characterizes the intrinsic spin dynamics of Anderson impurities in arbitrary conduction bands.

Recent NRG improvements of the full density-matrix (FDM) generalization \cite{Weichselbaum2007} based on the complete basis set of discarded states \cite{Anders2005} may fulfill this purpose. This FDM approach avoids the overcounting ambiguity and the single-shell approximation for the density matrix in the previous NRG, giving rise to very accurate dynamical correlation functions for local operators at arbitrary temperature. Within the FDM algorithm, the zeroth-moment spectral sum rule holds exactly to machine precision \cite{Weichselbaum2007,Anders2005}. Higher-moment spectral sum rules are also fulfilled with a high accuracy \cite{Zitko2009}. This motivates us to wonder whether the local susceptibility, as a ($-1$)th moment of the impurity-spin correlation function, can be calculated accurately from the FDM method, with an accuracy even higher than the impurity susceptibility. Note that the accuracy of the impurity susceptibility from the FDM, as indicated by Ref.\,[10], is only in the percent range. Besides, at low temperature, there also exists a severe artifact in the FDM algorithm to $\chi_\textrm{imp}$ (detailed in Sec.\,III A).

In this paper, we apply the FDM approach to dynamically calculate the local susceptibility of an Anderson impurity, without involving numerical differentiations. It is shown that within the FDM algorithm, the local susceptibility $\chi_{\textrm{loc}}$ can be calculated very accurately in all parameter regimes, while the thermodynamic calculation of the impurity susceptibility $\chi_{\textrm{imp}}$ encounters severe artifacts at low temperature. By comparison with exact Bethe ansatz results, the FDM results of the local susceptibility $\chi_{\textrm{loc}}$ are found accurate in the permil range. This accuracy is obtained at much reduced computational cost, and is at least one order higher than the accuracy of the impurity susceptibility. We then revisit the previous studies \cite{Hofstetter1999,Silva2006,Zhuravlev2011,Mastrogiuseppe2014}, by applying the FDM algorithm for the local susceptibility $\chi_{\textrm{loc}}$ to the case with arbitrary energy-dependent conduction bands. The revisit reaches a definitive answer: as long as the host density of states is not gapped at the Fermi energy, the low-energy excitations of an Anderson impurity in arbitrary energy-dependent bands are still a Fermi liquid and paramagnetic. The resultant Kondo screening of the impurity magnetic moment is qualitatively same as in a flat wide band. For such systems, the exotic behaviors previously observed \cite{Hofstetter1999,Silva2006,Zhuravlev2011,Mastrogiuseppe2014} in the impurity susceptibility do not represent the correct Kondo physics of the impurity. In particular, we demonstrate that the non-Fermi-liquid property and diamagnetism are spurious behaviors arising from the additional susceptibility of conduction electrons, which may be vulnerable to the NRG discretization error.

\section{II. model Hamiltonian and fdm approach to spin susceptibilities}
We consider the single-impurity Anderson model described by the Hamiltonian $H=H_\textrm{bath}+H_\textrm{imp}+H_\textrm{int}$,
\begin{eqnarray}
H_\textrm{bath}&=&\sum_{k,\sigma}\varepsilon_{k}C^\dagger_{k\sigma}C_{k\sigma},\\
H_\textrm{imp}&=&\sum_\sigma\varepsilon_{d}d^\dagger_\sigma d_\sigma+Un_\uparrow n_\downarrow,\\
H_\textrm{int}&=&\sum_{k,\sigma}V_kC^\dagger_{k\sigma}d_\sigma+\textrm{H.c.}.
\end{eqnarray}
where $C^\dagger_{k\sigma}$ ($d^\dagger_\sigma$) creates an electron with energy $\varepsilon_{k}$ ($\varepsilon_{d}$) and spin $\sigma=\uparrow,\downarrow$ in the bath (impurity), $n_\sigma=d^\dagger_\sigma d_\sigma$, $U$ parametrizes the on-site Coulomb repulsion, and the two subsystems are coupled via the hybridization $V_k$. The influence of the bath on the impurity's dynamics is fully determined by the hybridization function $\Gamma(\varepsilon)=\pi\sum_k|V_k|^2\delta(\varepsilon-\varepsilon_k)$, which depends on specific realizations of the electron bath. In the standard case, $\Gamma(\varepsilon)$ is constant for $|\varepsilon|\leq D$ and zero otherwise, with $D$ being the half bandwidth. While $D$ is usually the largest energy scale of the problem, the effect of finite bandwidth becomes important in the narrow-band model \cite{Hofstetter1999}. For a magnetic impurity adsorbed on the top of a carbon atom in graphene with Rashba spin-orbit interaction, the resulting hybridization function of the model has a linear energy dependence with sharp discontinuities \cite{Mastrogiuseppe2014}. $\Gamma(\varepsilon)$ may have even singularities around the Fermi energy, such as for magnetic impurities in square lattice \cite{Zhuravlev2011}. When the single-impurity Anderson model pertains to the double quantum-dot system consisting of an interacting dot coupled to the leads through another noninteracting dot, the impurity is coupled to an effective bath with a Lorentzian hybridization function \cite{Silva2006}.  Following, we calculate the local spin susceptibility $\chi_\textrm{loc}$ of the above Anderson model with arbitrary forms of $\Gamma(\varepsilon)$ by using the FDM-NRG method.

The NRG strategy \cite{NRG1980,Bulla2008} starts from discretizing the bath spectrum on a logarithmic grid of energies $\pm D\Lambda^{-n}$ with $\Lambda>1$ and $n=0,1,2,\cdots$, thereby transforming the original impurity model into a semi-infinite tight-binding chain with exponentially decreasing hopping matrix elements, via a standard tridiagonalization procedure after dropping high-mode states in each discretization interval. With its first site representing the impurity, the chain is then diagonalized iteratively adding one site at each step. In order to restrict the exponentially growing Hilbert space, eigenstates of the chain Hamiltonian $H_N$ including the newly added site (the $N$th site), are constructed from the states of the $N$th site and the $M_K$ lowest-lying eigenstates (kept states) of the chain $H_{N-1}$ without the $N$th site, while discarding the remaining eigenstates of $H_{N-1}$. The iterative diagonalization proceeds until the hopping matrix element between the last added site, say $N=N_\textrm{max}$, and its immediate neighbor becomes the smallest energy scale of the problem, such that the Hamiltonian $H_{N_\textrm{max}}$ of the full chain represents a good approximation of the original Anderson model.

Anders and Schiller \cite{Anders2005} have introduced a complete basis set of the Fock space of $H_{N_\textrm{max}}$ by constructing the tensor-product state $|l,e;N\rangle\equiv|l;N\rangle\otimes|\alpha_{N+1}\rangle\otimes|\alpha_{N+2}\rangle\cdots\otimes|\alpha_{N_\textrm{max}}\rangle$ from $|l;N\rangle$ the $l$th discarded states of $H_N$ and $|\alpha_m\rangle$ the state of the $m$th site with $\alpha_m=\{0,\uparrow,\downarrow,\uparrow\downarrow\}$, where $e$ denotes collectively the degrees of freedom of the sites $m=N+1,\cdots,N_\textrm{max}$, i.e., the environment of $H_N$. Let $N_\textrm{min}$ being the first iteration at which high-energy states are discarded and taking all eigenstates of the last iteration $N_\textrm{max}$ as discarded, the completeness relation of the basis set $\{|l,e;N\rangle\}$ reads as
\begin{equation}
\sum_{N=N_\textrm{min}}^{N_\textrm{max}}\sum_{l,e}|l,e;N\rangle\langle l,e;N|=1,
\end{equation}
along with the orthonormality
\begin{equation}
\langle l,e;N|l',e';N'\rangle=\delta_{ll'}\delta_{ee'}\delta_{NN'},
\end{equation}
and a useful identity for the subspaces spanned by the kept states (denoted with $k$) at iteration $N$, and by all discarded states at succedent iterations $N'>N$:
\begin{equation}
\sum_{k,e}|k,e;N\rangle\langle k,e,N|=\sum_{N'=N+1}^{N_\textrm{max}}\sum_{l,e}|l,e;N'\rangle\langle l,e;N'|.
\end{equation}
Note also that for $N'\leq N$, the kept and discarded states are orthogonal $\langle k,e;N|l,e';N'\rangle=0$. Since $|s,e;N\rangle$ ($s=l,k$) is only an exact eigenstate of $H_N$ corresponding to an eigenvalue $E^N_s$ with $4^{N_\textrm{max}-N}$-fold degeneracy, one has to assume it is also an eigenstate of the original model, $H|s,e;N\rangle\approx E^N_s|s,e;N\rangle$. This so-called NRG approximation represents the only approximation of the FDM algorithm. Weichselbaum and Delft \cite{Weichselbaum2007} hence write the full density matrix $\rho$ of $H$ as follows:
\begin{eqnarray}
\rho&=&\frac{1}{Z}\sum_{N=N_\textrm{min}}^{N_\textrm{max}}\sum_{l,e}e^{-\beta E^N_l}|l,e;N\rangle\langle l,e;N|,\\
Z&=&\sum_{N=N_\textrm{min}}^{N_\textrm{max}}\sum_{l}4^{N_\textrm{max}-N}e^{-\beta E^N_l},
\end{eqnarray}
with $\beta=1/(k_BT)$ the inverse temperature. By using this form of the density matrix, the complete basis set of discarded states, and the NRG approximation, all dynamic and static properties of $H$ can be evaluated.

By definition, the local susceptibility describes the impurity magnetization in response to a weak magnetic field $B$ applying only at the impurity site \cite{Bulla2008}
\begin{equation}
\chi_\textrm{loc}\equiv\lim_{B\rightarrow 0}\frac{\partial\langle M_\textrm{m}\rangle_{H+H'_m}}{\partial B}=(g\mu_B)^2\int^\beta_0\textrm{d}\tau\,\langle S_z(\tau)S_z\rangle_H,
\end{equation}
where $M_\textrm{m}=g\mu_B S_z$ is the impurity magnetization operator, $g$ the Land\'{e} $g$ factor, $\mu_B$ the Bohr magneton, $S_z=\frac{1}{2}(n_\uparrow-n_\downarrow)$ is the $z$ component of impurity spin, and $S_z(\tau)=e^{\tau H}S_ze^{-\tau H}$. $\langle\cdots\rangle_{H+H'_m}$ and $\langle\cdots\rangle_H$ denote the thermodynamic average with respect to the Hamiltonian $H$ with and without the perturbation $H'_m=-g\mu_BS_zB$. The second equality in Eq.\,(9), used $\langle S_z\rangle_H=0$, represents an exact mathematical relation which expresses a response in the impurity magnetization due to an infinitesimal local field in terms of an imaginary-time Matsubara Green's function \cite{Bulla2008}. This already constitutes an operational ground for dynamical calculating the local susceptibility, while obviating the need to evaluate a numerical derivative. However, the FDM algorithm for the Matsubara function would involve the NRG approximation, $e^{\pm\tau H}|s,e;N\rangle\approx e^{\pm \tau E^N_s}|s,e;N\rangle$, which at low temperature becomes severe \cite{note1} because of large $\tau$ involved.

To obtain high-quality susceptibility data in the low-temperature Kondo regime, it is thus better to work with the retarded Green's function for which the quality of the NRG approximation does not rely on the temperature. Using the Kubo formula for linear response in the static limit, the local susceptibility can be rewritten as
\begin{eqnarray}
\chi_\textrm{loc}&=&-(g\mu_B)^2\,\textrm{Re}[G_{S_z}(0)]\nonumber\\
&=&-(g\mu_B)^2\frac{1}{\pi}\,\mathcal{P}\int^{\infty}_{-\infty}\textrm{d}\varepsilon\,\frac{\textrm{Im}[G_{S_z}(\varepsilon)]}{\varepsilon},\\
G_{S_z}(\varepsilon)&=&\frac{1}{i\hbar}\int^\infty_{-\infty}\textrm{d}t\,e^{\frac{i}{\hbar}\varepsilon t}\Theta(t)\langle[S_z(t),S^\dagger_z(0)]\rangle_H.\quad
\end{eqnarray}
It is easy to verify the equivalence of Eqs.\,(9) and (10) from their Lehmann representations. Moreover, the second equality in Eq.\,(10) reveals that the local susceptibility is just the $(-1)$th moment of the impurity-spin spectral function. To evaluate within the FDM algorithm \cite{Weichselbaum2007} the retarded Green's function of the local spin, $G_{S_z}(\varepsilon)$, the full density matrix Eq.\,(7) and the complete basis set Eq.\,(4) are inserted into the thermal average in Eq.\,(11). Making use of the properties of the basis [Eqs.\,(5) and (6)] and taking the NRG approximation $e^{\pm\frac{i}{\hbar}Ht}|s,e;N\rangle\approx e^{\pm\frac{i}{\hbar}E^N_st}|s,e;N\rangle$, we end up with (see Appendix for details)
\begin{eqnarray}
\frac{\chi_\textrm{loc}}{\beta(g\mu_B)^2}&=&\hspace{-0.2cm}\sum_{N,N'=N_\textrm{min}}^{N_\textrm{max}}\hspace{-0.6cm}\,'\hspace{0.4cm}\sum_{k,k',l}
C^{(1)}_{Nkl}\cdot\rho^{NN'}_{k'k}\cdot[S_z]^N_{kl}\cdot[S_z]^N_{lk'}\nonumber\\
&&+\hspace{0.15cm}\sum_{N=N_\textrm{min}}^{N_\textrm{max}-1}\sum_{k,l}C^{(2)}_{Nkl}\cdot[S_z]^N_{kl}\cdot[S_z]^N_{lk}\nonumber\\
&&+\hspace{0.15cm}\sum_{N=N_\textrm{min}}^{N_\textrm{max}}\sum_{l,l'}C^{(3)}_{Nll'}
\cdot[S_z]^N_{ll'}\cdot[S_z]^N_{l'l}.\hspace{0.57cm}(12)\nonumber
\end{eqnarray}
Here $\sum'$ restricts the summation to $N<N'$, the matrix elements of local spin $[S_z]^N_{ss'}\equiv\langle s;N|S_z|s';N\rangle$, and the reduced density matrix $\rho^{NN'}_{k'k}\equiv\sum_e\langle k',e;N|\rho_{N'}|k,e;N\rangle$ in which the density matrix of shell $N'$,
\setcounter{equation}{12}
\begin{equation}
\rho_{N'}=\frac{1}{Z}\sum_{l,e}e^{-\beta E^{N'}_l}|l,e;N'\rangle\langle l,e;N'|,
\end{equation}
satisfies $\rho=\sum_{N=N_\textrm{min}}^{N_\textrm{max}}\rho_N$. Finally, the three coefficients in Eq.\,(12) are given by
\begin{eqnarray}
C^{(1)}_{Nkl}&=&\textrm{Re}\bigg[\frac{2}{\beta(E^N_l-E^N_k)+i\eta}\bigg],\\
C^{(2)}_{Nkl}&=&-\frac{4^{N_\textrm{max}-N}}{Z}e^{-\beta E^N_l}C^{(1)}_{Nkl},\\
C^{(3)}_{Nll'}&=&\frac{4^{N_\textrm{max}-N}}{Z}\frac{e^{-\beta E^N_{l'}}-e^{-\beta E^N_l}}{\beta(E^N_l-E^N_{l'})},
\end{eqnarray}
where $\eta$ is a dimensionless infinitesimal (already absorbed $\beta$) to deal with the accidental degeneracy of kept and discarded states, $E^N_k=E^N_l$, which appears when high-energy states are truncated at a degenerate eigenenergy of $H_N$. Keeping $\eta$ fixed here is equivalent to broaden discrete $\delta$ functions in the spectral function $\textrm{Im}[G_{S_z}(\omega)]$ using a Lorentzian kernel with temperature-dependent width $(\eta/\beta)$. In this work, we always set $\eta=0.001$, unless stated otherwise. Since the value of $C^{(3)}_{Nll'}$ in the limit $E^N_l=E^N_{l'}$ is well defined, there is no need to introduce an infinitesimal imaginary part to Eq.\,(16). After performing a ``forward run'' along the Wilson chain to iteratively generate all relevant NRG eigenenergies $E^N_s$, eigenstates $|s;N\rangle$, and matrix elements $[S_z]^N_{ss'}$, we can evaluate the reduced density matrix $\rho^{NN'}_{kk'}$ in a single ``backward run'' \cite{Weichselbaum2007}, thereby obtaining all information needed for dynamically calculating the local susceptibility.

For comparison, we also present the FDM results for the impurity contribution to susceptibility. This widely used quantity is defined originally by Wilson \cite{NRG1980,Bulla2008} to be the difference of the total magnetic response with and without the impurity, under a global field $B$,
\begin{eqnarray}
\chi_\textrm{imp}&\equiv&\lim_{B\rightarrow 0}\frac{\partial}{\partial B}\bigg[\langle M_\textrm{m}\rangle_{H+H'_t}+\langle M_\textrm{b}\rangle_{H+H'_t}\nonumber\\
& &\qquad\qquad\quad-\langle M_\textrm{b}\rangle_{H_\textrm{bath}+H'_b}\bigg]\nonumber\\
&=&(g\mu_B)^2\beta(\langle S_t^2\rangle_H-\langle S^2_b\rangle_{H_\textrm{bath}}),
\end{eqnarray}
where $H'_t=-g\mu_BS_tB$, $M_\textrm{b}=g\mu_BS_b$, $H'_b=-g\mu_BS_bB$, and $S_t=S_b+S_z$, with $S_b$ being the $z$ component of bath spin. The second equality, using $[S_t,\,H]=[S_b,\,H_\textrm{bath}]=0$, indicates that $\chi_\textrm{imp}$ is a thermodynamic quantity which, as usual, can be obtained accurately using the traditional NRG. Within the FDM method, at the $N$th iteration, both $S_{t}$ and $S_{b}$ should be divided into (i) the $z$-component spin, $S_N$, of the resultant chain Hamiltonian $H_N$ with or without the impurity site, and (ii) the $z$-component spin of $N_\textrm{max}-N$ environmental sites. Following Ref.\,[10], the FDM formula for $\chi_\textrm{imp}$ is
\begin{eqnarray}
&&\frac{\chi_\textrm{imp}}{\beta(g\mu_B)^2}=X^{\textrm{with impurity}}-X^{\textrm{without impurity}},\\
&&X=\frac{1}{Z}\sum_{N=N_\textrm{min}}^{N_\textrm{max}}
\sum_l4^{N_\textrm{max}-N}e^{-\beta E^N_l}\Big\{[S^2_{N}]^N_{ll}\qquad\ \nonumber\\
&&\qquad\qquad\qquad+\ \ (N_\textrm{max}-N)/8\Big\}.
\end{eqnarray}
At first glance, the absence of the reduced density matrix in Eq.\,(19) implies that thermodynamically evaluating $\chi_\textrm{imp}$ by two NRG runs (one with and one without the impurity) would require less computational resources than dynamically evaluating $\chi_\textrm{loc}$ by a ``forward'' and ``backward'' NRG run. As we show in the following, however, within the FDM approach the accuracy obtainable for $\chi_\textrm{loc}$ even at lower computational cost can be one order higher than that of $\chi_\textrm{imp}$.

To establish a relation between the local $\chi_\textrm{loc}$ and impurity $\chi_\textrm{imp}$ susceptibilities, let us consider their difference $\chi_\textrm{loc}-\chi_\textrm{imp}$. From the equation of motion (EOM) of the bath-electron Green's function, one readily obtains
\begin{eqnarray}
\chi_{\textrm{loc}}-\chi_{\textrm{imp}}&=&g\mu_B\lim_{B\rightarrow0}\frac{\partial}{\partial B}\sum_\sigma\sigma\hspace{-0.1cm}\int\frac{\textrm{d}\varepsilon}{2\pi}\,f(\varepsilon)\textrm{Im}\bigg[\delta G(\varepsilon)\nonumber\\
&&\qquad\qquad\quad-G_{d_\sigma}(\varepsilon)\frac{\partial\Sigma_\sigma(\varepsilon)}{\partial\varepsilon}\bigg],
\end{eqnarray}
where $f(\varepsilon)$ stands for the Fermi-Dirac function, $\Sigma_\sigma(\varepsilon)=\sum_k|V_k|^2/(\varepsilon-\varepsilon_k+\frac{1}{2}\sigma g\mu_BB+i0^+)$ is the self energy due to the impurity-bath coupling, and $\delta G(\varepsilon)=G_{d_\sigma}(\varepsilon)-G'_{d_\sigma}(\varepsilon)$, with $G_{d_\sigma}(\varepsilon)$ [$G'_{d_\sigma}(\varepsilon)$] being the impurity retarded Green's function corresponding to the Hamiltonian $H$ in the global (local) magnetic field. The term containing $G_{d_\sigma}(\varepsilon)\partial\Sigma_\sigma(\varepsilon)/\partial\varepsilon$ in Eq.\,(20) represents the additional susceptibility from the conduction electrons \cite{Merker2012}. For a flat band in the wide-band limit, one has exactly $\partial\Sigma_\sigma(\varepsilon)/\partial\varepsilon=0$ and $\delta G(\varepsilon)=0$, leading to $\chi_\textrm{loc}=\chi_\textrm{imp}$ as expected from the Clogston-Anderson compensation theorem \cite{Clogston1961}. On the other hand, the breakdown of this theorem in the presence of a narrow bandwidth and/or strong energy dependence in $\Sigma_\sigma(\varepsilon)$ [or $\Gamma(\varepsilon)$, equivalently] \cite{Santoro1991} can result in $\chi_\textrm{loc}$ and $\chi_\textrm{imp}$ differing substantially. In this case, the impurity susceptibility may be no longer suitable for faithfully characterizing the intrinsic spin dynamics of the impurity, especially when the additional bath susceptibility becomes significant.

\section{III. results and discussions}

What follows are the numerical results calculated by the FDM NRG in the units of $D=g\mu_B=k_B=1$. Technically, the NRG is still an approximation method that involves
the discretization error (controlled by the discretization parameter $\Lambda$) and the truncation error (controlled by the number $M_K$ of kept states in each iteration and also $\Lambda$) \cite{Zitko2011}. The two types of errors are interrelated. For coarser discretization at larger $\Lambda$, the discretization error increases, whereas the truncation error decreases due to the enhanced separation of energy scales. Large $\Lambda$ may also introduce spurious oscillations into thermodynamic quantities. These oscillations can be removed by using the z-averaging procedure, wherein one averages the final results from independent NRG calculations for $N_z$ interleaved discretization meshes $\pm D\Lambda^{-n+z(1-\delta_{n0})}$, with $N_z$ values of the twist parameter $z$ equally distributed in $[0,1)$. As for $M_K$, unlike in the conventional method, the number of kept states needed by the FDM algorithm can be largely reduced while still obtaining satisfactory accuracy for physical observables, due to the use of a complete basis set. Particularly, the sum-rule nature inherent in the FDM algorithm to $\chi_{\textrm{loc}}$ makes the local susceptibility very insensitive to $M_K$. We thus always reduce the number of kept states for calculating $\chi_{\textrm{loc}}$ to half the number of states kept for $\chi_{\textrm{imp}}$. Nevertheless, one still has to carefully choose these parameters in any practical NRG calculations such that the resultant discretization and truncation errors do not affect physical conclusions drawn from the NRG data, which should be robust to changes of parameters.

\subsection{A. Impurities in the flat wide conduction band}
We first examine the FDM-NRG calculations of the two susceptibilities for a flat conduction band [$\Gamma(\varepsilon)=\Gamma$] in the wide-band limit ($D\gg\Gamma, U, |\varepsilon_d|$). The calculations are performed for the temperature dependence of the susceptibilities, by fixing $\Gamma$ while varying the on-site Coulomb repulsion and the impurity level position, as shown in Figs.\,1 and 2 and Tables I and II. Even though the two susceptibilities are essentially identical in this case, one can not expect the FDM-NRG calculation would always yield equal results for them. This is because the errors inherent in NRG may affect $\chi_{\textrm{loc}}$ and $\chi_{\textrm{imp}}$ in different ways since different algorithms as Eqs.\,(12) and (18) are adopted. For small $\Lambda$ and moderate $M_K$ [Fig.\,1(a)], it is shown that at high temperatures both algorithms have introduced only tiny artifacts. But a severe artifact appears in the low-temperature behavior of $\chi_{\textrm{imp}}$, which obviously breaks the Fermi-liquid property. At low temperatures, the Fermi-liquid theory \cite{Hewson1993} of a Kondo impurity demonstrates a linear temperature dependence of the effective Curie constant $T\chi$, i.e., the spin susceptibility $\chi$ would develop a plateau as $T\rightarrow0$. This plateau is indeed very well formed in $\chi_{\textrm{loc}}$ as a function of $T$, indicating that dynamical FDM calculation of the local susceptibility is much more accurate than the thermodynamic calculation of the impurity susceptibility.

\begin{figure}[!htb]
\centering
\includegraphics[width=1.0\columnwidth]{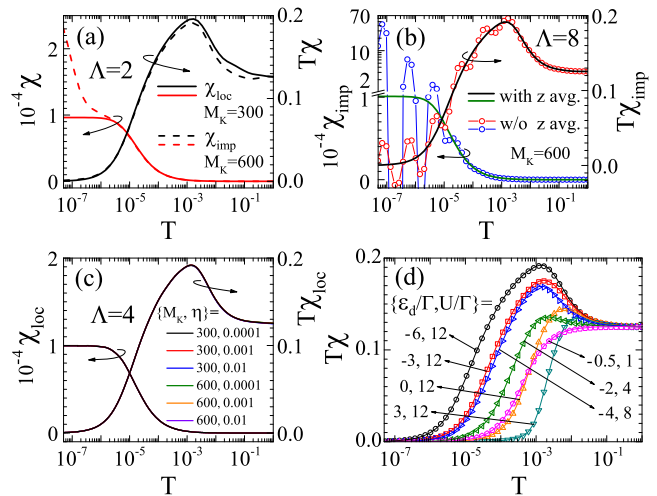}
\caption{(Color online) Local $\chi_{\textrm{loc}}$ and impurity $\chi_{\textrm{imp}}$ susceptibilities calculated by the FDM NRG for the Anderson model with a flat wide band, over a wide temperature range continuously evolving from the high-temperature free-orbital to the low-temperature Fermi-liquid regimes. Model parameters: $D=1$, $\Gamma=0.001$, $U/\Gamma=12$, and $\varepsilon_d/\Gamma=-6$, unless indicated otherwise. Note that the curves in (c) regarding different parameters are highly coincident. NRG parameters in (d): for $\chi_{\textrm{loc}}$ (lines), $\Lambda=4$, $M_K=300$; for $\chi_{\textrm{imp}}$ (symbols), $\Lambda=8$, $M_K=600$. The z averaging is performed only for $\chi_{\textrm{imp}}$ in (b) and (d), using $N_z=8, z=0, 0.125, 0.25, 0.375, 0.5, 0.625, 0.75, 0.875$.}
\end{figure}

The artifacts uncovered in Fig.\,1(a) are mostly due to the truncation errors which can be largely reduced by increasing the discretization parameter $\Lambda$. For the impurity susceptibility $\chi_{\textrm{imp}}$, we find that using a discretization parameter as large as $\Lambda=8$ and performing the z averaging for $N_z=8$ meshes are necessary in order to obtain satisfactory data [Fig.\,1(b)], in agreement with Ref.\,[10]. On the other hand, high-quality data of $\chi_{\textrm{loc}}$, robust to changes of the number of kept states $M_K$ and the parameter $\eta$, are already available at moderate $\Lambda=4$, as given in Fig.\,1(c). Since at this value of $\Lambda$ the local susceptibility $\chi_{\textrm{loc}}$ exhibits no spurious oscillations, there is no need to carry out the z averaging. This significantly lowers the computational cost. With these NRG parameters tuned independently for $\chi_{\textrm{loc}}$ and $\chi_{\textrm{imp}}$ to eliminate the artifacts, the two susceptibilities plotted as $T\chi$ vs $T$ in Fig.\,1(d) seem indeed identical in all parameter regimes. The resultant curves exhibit features of typical Kondo screening for $-\varepsilon_d,\,\varepsilon_d+U\gg\Gamma$, that is, $T\chi$ first increasing from the high-temperature value $1/8$ of a free impurity towards its local-moment value $1/4$ for intermediate temperature, and then falling to zero as $T\rightarrow0$ due to the screening of the local magnetic moment by conduction electrons.

However, a severe drawback of the FDM method to $\chi_{\textrm{imp}}$, obscured by the very small values of $T\chi_{\textrm{imp}}$ at low temperature and thus overlooked by Ref.\,[10], is that the low-temperature artifacts appearing in $\chi_{\textrm{imp}}$ can not be completely eliminated in all parameter regimes, even though large $\Lambda$ and the z averaging are used. To explicitly demonstrate this, we present by Fig.\,2 a detailed comparison of the two susceptibilities, plotted as $\chi$ vs $T$ rather than $T\chi$ vs $T$, emphasizing the low-temperature Fermi-liquid plateau. As depicted, our FDM calculation of the local susceptibility gives high-quality data for all impurity parameters, ranging from the strongly correlated to the noninteracting regime [Fig.\,2(a)], and from the Kondo to the mixed valence and into the empty-orbital regime [Fig.\,2(b)], whereas the data of the impurity susceptibility become wildly irregular at extremely low temperature. We have checked that such kind of irregularities always show up in $\chi_{\textrm{imp}}$ by further increasing $\Lambda$, $N_z$, and even $M_K$. Seemingly, these irregularities are small in the Kondo regime and in the particle-hole symmetry case, but becomes severe in the mixed-valence and empty-orbital regimes. The precise origin of these irregularities is not clear at present. It may be \cite{Costi2015} due to an imperfect cancellation of the logarithmic discretization oscillations by the simple z-averaging procedure, or the limitations in numerical precision for calculating $\chi_{\textrm{imp}}$ by taking the subtraction of two extensive macroscopic values to obtain an impurity-related finite quantity. In any case, the FDM gives indisputably better data quality for the local susceptibility than for the impurity susceptibility, being distinct from the conventional NRG.

\begin{figure}[]
\includegraphics[width=0.9\columnwidth]{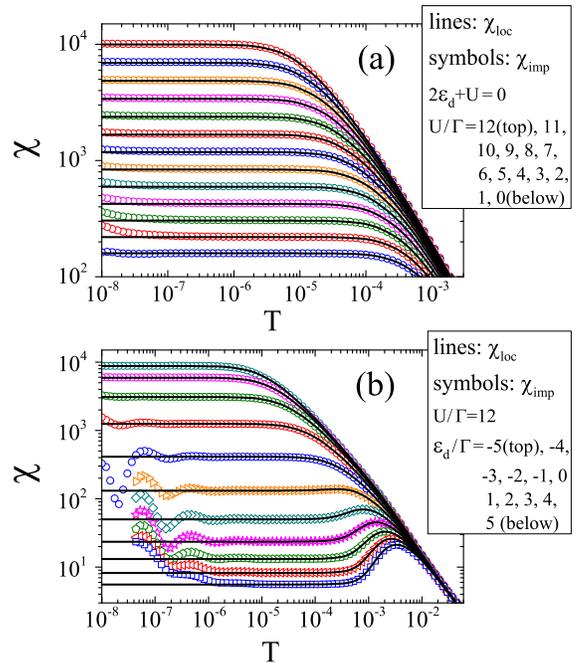}
\caption{(Color online) FDM-NRG results of the local and impurity susceptibilities plotted as $\chi$ vs $T$, focusing on the low-temperature Fermi-liquid plateau, for the symmetric (a) and asymmetric (b) Anderson models with a flat wide band ($D=1$, $\Gamma=0.001$). NRG and z-averaging parameters are the same as in Fig.\,1(d).}
\end{figure}

To quantitatively demonstrate the advantage of our FDM algorithm for the local susceptibility, we compare both susceptibilities with the exact zero-temperature results $\chi_{\textrm{BA}}$ from the Bethe ansatz \cite{BA1983}, as shown in Tables I and II. Note that one can not take exactly $T=0$ in FDM-NRG calculations. The zero-temperature results within NRG are actually extracted from the corresponding low-temperature results which must be convergent as $T\rightarrow0$. This poses no difficulties for the local susceptibility, because $\chi_{\textrm{loc}}$ do indeed converge to a definite value as $T\rightarrow0$ (see Fig.\,2). We thus take the zero-temperature value of $\chi_{\textrm{loc}}$ at the lowest temperature involved in our calculation, i.e., $T=10^{-8}$. However, due to its low-temperature artifacts, extracting the zero-temperature value of $\chi_{\textrm{imp}}$ would be problematic since $\chi_{\textrm{imp}}(T)$ is not convergent as $T\rightarrow0$. To avoid the problem, the zero-temperature impurity susceptibility we used in Tables I and II for comparison is actually the values of $\chi_{\textrm{imp}}(T)$ in the plateau region at $T=10^{-3}T_0$, where $T_0$ is the low-energy Kondo scale defined by $T_0\chi_{\textrm{imp}}(T_0)=0.0701$ \cite{NRG1980}. Comparison with the Bethe ansatz results indicates a relative error of $\chi_{\textrm{loc}}$ from the FDM in the permil range, while $\chi_{\textrm{imp}}$ is accurate only to within a few percent. This much improved accuracy of $\chi_{\textrm{loc}}$ becomes even more remarkable when considering the following fact. Our dynamical calculation of the local susceptibility does not need to perform the z-averaging, and includes only half of the eigenstates as kept in thermodynamically evaluating $\chi_{\textrm{imp}}$, and thus is carried out at much reduced computational cost.

\begin{table}[!t]
\renewcommand{\arraystretch}{1.2}
\caption{Zero-temperature local $\chi_{\textrm{loc}}$ and impurity $\chi_{\textrm{imp}}$ susceptibilities from the FDM NRG in comparison with the exact result $\chi_{\textrm{BA}}$ from the Bethe ansatz, for the symmetric Anderson model with a flat wide band [$D=1$(NRG), $\infty$(BA), $\Gamma=0.001$, $\varepsilon_d=-U/2$]. NRG and z-averaging parameters are the same as in Fig.\,1(d).}
\begin{tabular*}{1.0\columnwidth}{m{0.095\columnwidth}<{\centering}|m{0.16\columnwidth}<{\centering}
|m{0.16\columnwidth}<{\centering}m{0.16\columnwidth}<{\centering}|m{0.16\columnwidth}<{\centering}
m{0.16\columnwidth}<{\centering}}
\hline
\hline
\rule{0pt}{15pt}\raisebox{1mm}{$U/\Gamma$} & $\frac{\chi_{_\textrm{BA}}}{(g\mu_B)^2}$ & $\frac{\chi_{_\textrm{loc}}}{(g\mu_B)^2}$ & $\%$ error & $\frac{\chi_{_\textrm{imp}}}{(g\mu_B)^2}$ & $\%$ error\\
\hline
$12$ & $9970.8$ & $9974.0$ & $+0.03\%$ & $10136$ &  $\ +1.7\%$ \\
$11$ & $6950.4$ & $6943.9$ & $-0.09\%$ & $7063.7$ & $\ +1.6\%$ \\
$10$ & $4854.3$ & $4854.8$ & $+0.01\%$ & $4931.4$ & $\ +1.6\%$ \\
$9$  & $3397.7$ & $3398.3$ & $+0.02\%$ & $3451.3$ & $\ +1.6\%$ \\
$8$  & $2383.8$ & $2382.7$ & $-0.05\%$ & $2420.7$ & $\ +1.5\%$ \\
$7$  & $1676.9$ & $1673.9$ & $-0.18\%$ & $1702.5$ & $\ +1.5\%$ \\
$6$  & $1183.2$ & $1182.3$ & $-0.08\%$ & $1201.6$ & $\ +1.6\%$ \\
$5$  & $837.65$ & $837.13$ & $-0.06\%$ & $849.94$ & $\ +1.5\%$ \\
$4$  & $595.23$ & $594.40$ & $-0.14\%$ & $603.39$ & $\ +1.4\%$ \\
$3$  & $424.76$ & $423.41$ & $-0.32\%$ & $430.15$ & $\ +1.3\%$ \\
$2$  & $304.54$ & $304.01$ & $-0.17\%$ & $308.14$ & $\ +1.2\%$ \\
$1$  & $219.50$ & $219.68$ & $+0.08\%$ & $221.54$ & $\ +0.9\%$ \\
$0$  & $159.15$ & $159.09$ & $-0.04\%$ & $159.34$ & $\ +0.1\%$ \\
\hline
\hline
\end{tabular*}
\end{table}

\begin{table}[!t]
\renewcommand{\arraystretch}{1.2}
\caption{Zero-temperature local $\chi_{\textrm{loc}}$ and impurity $\chi_{\textrm{imp}}$ susceptibilities from the FDM NRG in comparison with the exact result $\chi_{\textrm{BA}}$ from the Bethe ansatz, for the asymmetric Anderson model with a flat wide band [$D=1$(NRG), $\infty$(BA), $\Gamma=0.001$, $U/\Gamma=12$]. NRG and z-averaging parameters are the same as in Fig.\,1(d).}
\begin{tabular*}{1.0\columnwidth}{m{0.095\columnwidth}<{\centering}|m{0.16\columnwidth}<{\centering}
|m{0.16\columnwidth}<{\centering}m{0.16\columnwidth}<{\centering}|m{0.16\columnwidth}<{\centering}
m{0.16\columnwidth}<{\centering}}
\hline
\hline
\rule{0pt}{15pt}\raisebox{1mm}{$\varepsilon_d/\Gamma$} & $\frac{\chi_{_\textrm{BA}}}{(g\mu_B)^2}$ & $\frac{\chi_{_\textrm{loc}}}{(g\mu_B)^2}$ & $\%$ error & $\frac{\chi_{_\textrm{imp}}}{(g\mu_B)^2}$ & $\%$ error\\
\hline
$-5$ & $8748.3$ & $8749.9$ & $+0.02\%$ & $8811.6$ & $\ +0.7\%$ \\
$-4$ & $5910.2$ & $5904.8$ & $-0.09\%$ & $5974.5$ & $\ +1.1\%$ \\
$-3$ & $3078.5$ & $3075.9$ & $-0.08\%$ & $3108.2$ & $\ +1.0\%$ \\
$-2$ & $1246.7$ & $1244.9$ & $-0.14\%$ & $1285.5$ & $\ +3.1\%$ \\
$-1$ & $412.21$ & $412.13$ & $-0.02\%$ & $391.54$ & $\ -5.0\%$ \\
$0$  & $131.75$ & $131.42$ & $-0.25\%$ & $140.54$ & $\ +6.7\%$ \\
$1$  & $50.045$ & $49.996$ & $-0.10\%$ & $49.232$ & $\ -1.6\%$ \\
$2$  & $23.607$ & $23.563$ & $-0.19\%$ & $23.084$ & $\ -2.2\%$ \\
$3$  & $13.150$ & $13.128$ & $-0.17\%$ & $13.294$ & $\ +1.1\%$ \\
$4$  & $8.2236$ & $8.2098$ & $-0.17\%$ & $8.5423$ & $\ +3.9\%$ \\
$5$  & $5.5769$ & $5.5774$ & $+0.01\%$ & $5.8513$ & $\ +4.9\%$ \\
\hline
\hline
\end{tabular*}
\end{table}

\subsection{B. Impurities in energy-dependent bands}
We now turn to clarify the effect of energy-dependent conduction bands on the magnetic response of an Anderson impurity. The single-impurity Anderson model with a structured conduction band has already been intensively studied in the literature. It has been shown that when there is a gap (hard or soft) at the Fermi energy in the host density of states, the impurity ground state can undergo quantum phase transitions from the Kondo screening to the local moment state, giving rise to non-Fermi-liquid behavior \cite{Fang2013}. However, there is also a class of systems \cite{Mastrogiuseppe2014,Zhuravlev2011,Silva2006,Hofstetter1999} in which the host density of states, although strongly energy-dependent, has no gap at the Fermi energy. Previous investigations \cite{Mastrogiuseppe2014,Zhuravlev2011,Silva2006,Hofstetter1999} of such systems suggest that the low-energy excitations of the impurity can still be a non-Fermi liquid and even diamagnetic. The observations are quite surprising and are based on the results of the impurity susceptibility $\chi_{\textrm{imp}}$ calculated by the traditional NRG. Since $\chi_{\textrm{imp}}$ can not directly reflect the magnetic states of the impurity, we thus revisit this problem by directly calculating the local susceptibility $\chi_{\textrm{loc}}$ of these systems using the FDM NRG, in order to clarify whether an energy-dependent band without gap could indeed render the impurity diamagnetic and a non-Fermi liquid, or the previous observation \cite{Mastrogiuseppe2014,Zhuravlev2011,Silva2006,Hofstetter1999} is just spurious.

The first system \cite{Mastrogiuseppe2014} we shall examine is a magnetic impurity adsorbed in graphene with Rashba spin-orbit interaction, in which the interplay of the Rashba coupling and the linear graphene dispersion results in an effective host density of states described by the following hybridization function \cite{Mastrogiuseppe2014}
\begin{equation}
\Gamma(\varepsilon)=\Gamma_0[|\varepsilon|+\lambda+(|\varepsilon|-\lambda)\Theta(|\varepsilon|-2\lambda)]/D,
\end{equation}
where $\lambda$ characterizes the magnitude of the Rashba interaction, the prefactor $\Gamma_0=\Omega_0DV^2/(4v^2_F)$ with $\Omega_0$ the graphene unit-cell area, $V$ the overlap between the impurity level and the nearest carbon $p_z$ orbital, and $v_F$ the Fermi velocity. This hybridization function has a linear energy dependence with discontinuities at $\varepsilon=\pm2\lambda$. When the Fermi energy $\mu$ of this system is tuned to lie exactly at the discontinuity $\mu=2\lambda$, Ref.\,[11] has found by the traditional NRG that $\lim_{T\rightarrow0}T\chi_{\textrm{imp}}$ is not zero but rather negative, i.e., $\lim_{T\rightarrow0}\chi_{\textrm{imp}}$ is negatively divergent by $1/T$. This implies a non-Fermi-liquid and diamagnetic behavior in the impurity ground state. For the Fermi energy not exactly at (but very close to) the discontinuity, while the Fermi-liquid property $\lim_{T\rightarrow0}T\chi_{\textrm{imp}}=0$ is restored, the impurity still passes through a temperature window of diamagnetic behavior ($\chi_{\textrm{imp}}<0$) \cite{Mastrogiuseppe2014}. We have calculated by the FDM NRG the local and impurity susceptibilities of this system. Results are presented in Fig.\,3. Note that our FDM results of $\chi_{\textrm{imp}}$ [Figs.\,3(c) and 3(d)] do indeed verify the results of Ref.\,[11], despite the fact that the FDM algorithm for $\chi_{\textrm{imp}}$ introduces artifacts at extremely low temperature. As explained already in Section III.A, these artifacts are invisible when plotted as $T\chi_{\textrm{imp}}$ vs $T$ [Fig.\,3(c)], but apparently show up in $\chi_{\textrm{imp}}$ vs $T$ [see the inset of Fig.\,3(d), where the artifacts have violated the Fermi-liquid plateau at $T<10^{-10}$ in the curves corresponding to $\mu=2\lambda\pm10^{-9}$].

However, the above non-Fermi-liquid and diamagnetic behavior found in Ref.\,[11] are not supported by our FDM results of the local susceptibility $\chi_{\textrm{loc}}$. Figures 3(a) and 3(b) demonstrate that for different values of the Rashba parameter, the low-energy excitations of the impurity are always a Fermi liquid ($\lim_{T\rightarrow0}T\chi_{\textrm{loc}}=0$, $\lim_{T\rightarrow0}\chi_{\textrm{loc}}=const.$) and paramagnetic ($\chi_{\textrm{loc}}>0$), no matter the Fermi level lies exactly at ($\mu=2\lambda$) or slightly deviates from ($\mu=2\lambda\pm10^{-9}$) the discontinuity. Moreover, unlike the behavior of $\chi_{\textrm{imp}}$, there is no significant difference in $\chi_{\textrm{loc}}$ for these positions of the Fermi level (see the red solid lines and the symbols in Fig.\,3). This behavior of $\chi_{\textrm{loc}}$ is consistent with the underlying Kondo physics, while $\chi_{\textrm{imp}}$ is not. Generally speaking, at temperatures much lower than the Kondo scale $T\ll T_0$, only those conduction electrons within the energy window $|\varepsilon-\mu|<T_0$ participate in the Kondo screening. Since the Fermi energies $\mu=2\lambda, 2\lambda\pm10^{-9}$ produce the almost same Kondo scale \cite{Mastrogiuseppe2014} which is far, far larger than the energy difference in these $\mu$, the same portion of conduction electrons around the Fermi level are involved in screening the impurity spin. Consequently, for these Fermi energies, the impurity magnetic response should also be almost equal, as indicated by our $\chi_{\textrm{loc}}$, rather than $\chi_{\textrm{imp}}$ in Ref.\,[11] suggested. We thus argue that the correct magnetic property of the impurity in graphene with the exotic hybridization function Eq.\,(21) is still a standard Fermi liquid.

\begin{figure}[]
\includegraphics[width=1.0\columnwidth]{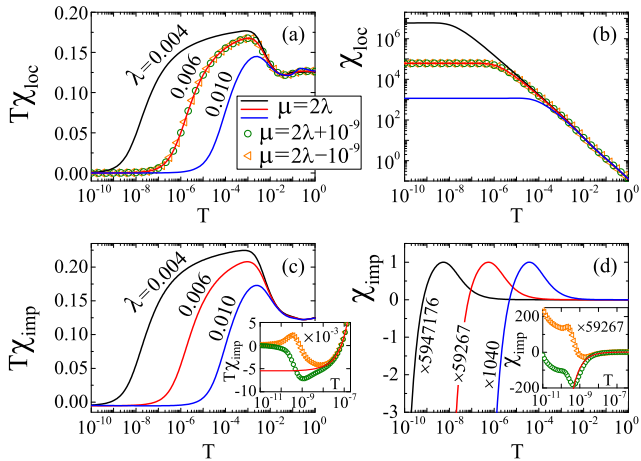}
\caption{(Color online) Local [(a),(b)] and impurity [(c),(d)] susceptibilities, plotted as $T\chi$ vs $T$ [(a),(c)] and $\chi$ vs $T$ [(b),(d)], from the FDM for a magnetic impurity in graphene with different Rashba spin-orbit coupling $\lambda$ and chemical potential $\mu$. The resulting Anderson model \cite{Mastrogiuseppe2014} is characterized by the hybridization function Eq.\,(21). The legend in (a) is applied to all figures including the insets. Insets: impurity susceptibility for different $\mu$ at fixed $\lambda=0.006$, plotted as $T\chi_{\textrm{imp}}$ vs $T$ [inset of (c)] and $\chi_{\textrm{imp}}$ vs $T$ [inset of (d)]. Model parameters: $U=-2\varepsilon_d=0.5\Gamma_0=0.02$, being the same as in Figs.\,5 and 7 of Ref.\,[11]. NRG parameters: $\Lambda=2.5$; for $\chi_{\textrm{loc}}$, $M_K=600$, without the z averaging; for $\chi_{\textrm{imp}}$, $M_K=1200$, $N_z=2$ with $z=0, 0.5$.}
\end{figure}

The second system \cite{Zhuravlev2011} we have revisited is a magnetic impurity in the two-dimensional square lattice, with the half bandwidth $D$ determined by the nearest-neighbor hopping energy $D=4t$. Its host density of states $\rho(\varepsilon)$ has a Van Hove singularity near the Fermi energy $\mu=0$. According to Ref.\,[12], the distance $\Delta$ from the singularity to the Fermi energy is $\Delta=4t'$ with $t'$ the next nearest-neighbor hopping, and
\begin{equation}
\rho(\varepsilon)=\frac{2\ln[(4\sqrt{D^2-\Delta^2})/(|\varepsilon+\Delta|)]}{\pi^2D\sqrt{1-(\Delta/D)^2}}.
\end{equation}
Based on the NRG results of $\chi_{\textrm{imp}}$ within the Kondo model, Ref.\,[12] predicted a non-Fermi-liquid and diamagnetic regime at low temperature for $\Delta=0$, in which $T\chi_{\textrm{imp}}\approx-0.072/|\ln(T/D)|^{0.77}$ (also leading to divergent $\lim_{T\rightarrow0}\chi_\textrm{imp}$). For nonzero but very small $\Delta$, the Fermi-liquid behavior $T\chi_{\textrm{imp}}=cT$ is restored with the scale factor $c$ remaining negative. These predictions are qualitatively verified by our FDM results of $\chi_{\textrm{imp}}$ based on the Anderson model, as shown in Fig.\,4(a). Our FDM results of $\chi_{\textrm{loc}}$ [also presented in Fig.\,4(a)] again do not support these non-Fermi-liquid and diamagnetic behaviors. While the authors of Ref.\,[12] attributed these spurious behaviors to an overcompensation of the local spin by the conduction electrons, we draw a conclusion from the local susceptibility that a Kondo impurity in the square lattice with Van Hove singularities is still a Fermi liquid and there is no diamagnetic or overscreening effects. Additionally, for the values of $\Delta$ used in plotting Fig.\,4(a), there are no sizable deviations in the local susceptibility. This is due to the same reason as already explained in the first system, i.e., these values are very close, which give rise to the nearly same Kondo temperature and thus involve the same portion of conduction electrons in the Kondo screening.

\begin{figure}[]
\includegraphics[width=0.8\columnwidth]{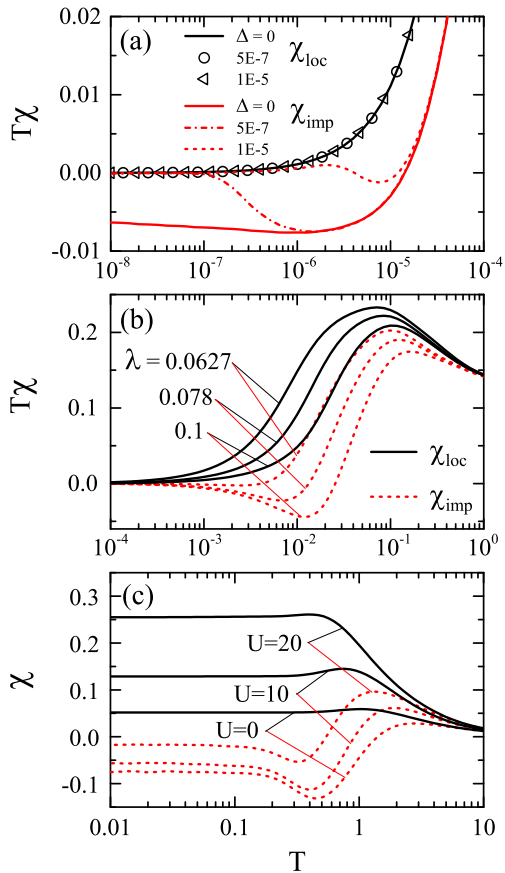}
\caption{(Color online) Local and impurity susceptibilities from the FDM for (a) a magnetic impurity in square lattice \cite{Zhuravlev2011}, (b) the double quantum dot system \cite{Silva2006}, and (c) a narrow-band system \cite{Hofstetter1999}. The resulting hybridization function in the Anderson model is (a) $\Gamma(\varepsilon)=\pi V^2\rho(\varepsilon)$ with $\rho(\varepsilon)$ given by Eq.\,(22), (b) given by Eq.\,(23), and (c) $\Gamma(\varepsilon)=\Gamma\Theta(D-|\varepsilon|)$ for $D\ll U,|\varepsilon_d|$. Model parameters: (a) $U=-2\varepsilon_d=0.05$, $2V^2/(\pi D)=0.0006$; being consistent with the Kondo-model parameters used by Ref.\,[12]; (b) $U=-2\varepsilon_d=0.5$, $\Delta_2=0.02$, being the same as in Fig.\,2(d) of Ref.\,[13]; (c) $U=-2\varepsilon_d$, $\Gamma=10$, being the same as in Fig.\,1 of Ref.\,[14]. NRG parameters: $\Lambda=2$; for $\chi_{\textrm{loc}}$, $M_K=600$; for $\chi_{\textrm{imp}}$, $M_K=1200$; without the z averaging.}
\end{figure}

We have also examined the effects of a Lorentzian and a narrow conduction band on the impurity spin dynamics. The Lorentzian host density of states can be realized by double quantum dot systems, where an interacting quantum dot (the impurity) is indirectly coupled to the leads through a noninteracting quantum dot \cite{Silva2006}.
When the resonance in the noninteracting dot is at the Fermi energy, the double quantum dot system maps onto a single-impurity Anderson model having a Lorentzian hybridization function \cite{Silva2006}
\begin{equation}
\Gamma(\varepsilon)=\frac{\lambda^2}{\Delta_2}\frac{\Delta_2^2}{\varepsilon^2+\Delta_2^2},
\end{equation}
where $\lambda$ is the interdot coupling and $\Delta_2$ the coupling between the noninteracting dot and the leads. On the other hand, the narrow-band Anderson model is generated by the dynamical mean field theory for the Mott-Hubbard transition in infinite dimensions \cite{Hofstetter1999}. Although the two models are of interest in such different contexts, the resulting impurity properties are similar since the Lorentzian band is in analogy with the narrow band in the sense that it diminishes the high-energy conduction states reducing effectively the bandwidth. Previous studies \cite{Silva2006,Hofstetter1999} within the traditional NRG show that for both models the impurity susceptibility suggests a Fermi liquid ($\lim_{T\rightarrow0}T\chi_{\textrm{imp}}=0$, $\lim_{T\rightarrow0}\chi_{\textrm{imp}}=const.$) in all parameter regimes, but also a diamagnetic ($\chi_{\textrm{imp}}<0$) region in $\chi_{\textrm{imp}}$ vs $T$ for some parameters. As expected, our $\chi_{\textrm{imp}}$ by the FDM repeats these properties [see the dashed curves in Figs.\,4(b) and 4(c)]. The Fermi-liquid property is also confirmed by our local susceptibility $\chi_{\textrm{loc}}$ presented in Figs.\,4(b) and 4(c). However, $\chi_{\textrm{loc}}$ for both models is always positive even in the parameter regimes where $\chi_{\textrm{imp}}$ is negative. This rules out the scenario of diamagnetic impurities caused by the Lorentzian or narrow conduction bands.

\subsection{C. Origins of the spurious diamagnetic and non-Fermi-liquid behaviors in $\chi_\textrm{imp}$}
To pinpoint the origin of the diamagnetism and non-Fermi-liquid behavior in the impurity susceptibility, let us look back into the definition Eq.\,(17) of $\chi_\textrm{imp}$. The first term $\textrm{lim}_{B\rightarrow 0}\frac{\partial}{\partial B}\langle M_\textrm{m}\rangle_{H+H'_t}$ in Eq.\,(17) is always positive and its contribution is very similar to the local susceptibility $\chi_\textrm{loc}$. Their difference gives rise to the $\delta G(\varepsilon)$ term in Eq.\,(20). The other two terms in Eq.\,(17), $\textrm{lim}_{B\rightarrow 0}\frac{\partial}{\partial B}(\langle M_\textrm{b}\rangle_{H+H'_t}-\langle M_\textrm{b}\rangle_{H_\textrm{bath}+H'_b})\equiv\delta\chi_c$, represents the additional susceptibility of conduction electrons induced by the presence of the impurity. $\delta\chi_c$ is exactly the $G_{d_\sigma}(\varepsilon)\partial\Sigma_\sigma(\varepsilon)/\partial\varepsilon$ term in the right-hand side of Eq.\,(20). Due to the derivative of the self energy, the sign and magnitude of $\delta\chi_c$ are very sensitive to the shape of the conduction band. It is this additional bath susceptibility included in the definition of $\chi_{\textrm{imp}}$ that becomes negative and divergent at low temperature when the host density of states is strongly energy dependent. Therefore, the diamagnetism and non-Fermi-liquid behavior found in previous studies [11-14] does not directly reflect the intrinsic impurity properties, and has nothing to do with the Kondo screening of the local moment at low temperature.

For a deep insight into the non-Fermi-liquid property suggested by $\chi_\textrm{imp}$ in the graphene \cite{Mastrogiuseppe2014} and the square lattice \cite{Zhuravlev2011} systems, we consider the $U=0$ Anderson model which allows to clarify wether the $\chi_{\textrm{imp}}\sim T$ dependence in the $U\neq0$ case is qualitatively different from the $U=0$ case and also allows to compare the FDM-NRG results with the exact ones. In the noninteracting case, the exact local and impurity susceptibilities can be obtained using the EOM approach,
\begin{eqnarray}
\frac{\chi_\text{loc}}{(g\mu_B)^2}&=&\int\frac{\textrm{d}\varepsilon}{2\pi}f(\varepsilon)\textrm{Im}[\varepsilon-\varepsilon_d-\Sigma(\varepsilon)]^{-2},\\
\frac{\chi_\text{imp}}{(g\mu_B)^2}&=&\int\frac{\textrm{d}\varepsilon}{2\pi}f(\varepsilon)\textrm{Im} \Bigg\{\frac{\frac{\partial^2}{\partial\varepsilon^2}\Sigma(\varepsilon)}{\varepsilon-\varepsilon_d-\Sigma(\varepsilon)}\nonumber\\
&&\qquad\qquad\quad+\bigg[\frac{1-\frac{\partial}{\partial\varepsilon}\Sigma(\varepsilon)}{\varepsilon-\varepsilon_d-\Sigma(\varepsilon)}\bigg]^2\Bigg\},\\
\Sigma(\varepsilon)&=&\frac{1}{\pi}\int\textrm{d}\varepsilon'\,\frac{\Gamma(\varepsilon')}{\varepsilon-\varepsilon'+i0^+}.
\end{eqnarray}
Note that for arbitrary energy-dependent bands ungapped at the Fermi energy, Eq.\,(24) always give rise to finite (not divergent) values of $\chi_\textrm{loc}$, as the temperature $T\rightarrow0$. But this is not the case for Eq.\,(25) of $\chi_{\textrm{imp}}$. Figure 5 presents the FDM-NRG results for the spin susceptibilities of the noninteracting Anderson impurity adsorbed in graphene [Fig.\,5(a)] with the hybridization function Eq.\,(21) and in the square lattice [Fig.\,5(b)] with the density of states Eq.\,(22). These are in good agreement with the exact EOM results. It is demonstrated by Fig.\,5 that even in the $U=0$ case, the impurity susceptibility $\chi_{\textrm{imp}}$ is already negatively divergent as $T\rightarrow0$, being qualitatively analogous to the corresponding interacting systems [see Figs.\,3(c), 3(d), and 4(a)]. This qualitative analogy between the interacting and noninteracting systems confirms again that the graphene \cite{Mastrogiuseppe2014} and square lattice \cite{Zhuravlev2011} systems are indeed a Fermi liquid \cite{Hewson1993}. From this point of view, the non-Fermi-liquid physics proposed previously \cite{Zhuravlev2011} according to the divergence of $\chi_{\textrm{imp}}$ as $T\rightarrow0$ is conceptually incorrect and misleading.

\begin{figure}[]
\includegraphics[width=0.8\columnwidth]{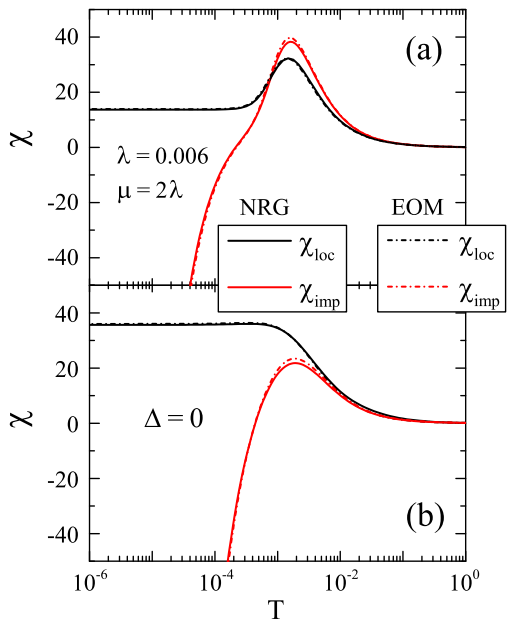}
\caption{(Color online) Local and impurity susceptibilities from the FDM, for a noninteracting impurity (a) in graphene with the hybridization function Eq.\,(21) and (b) in the square lattice with the density of states Eq.\,(22). Model parameters: (a) $U=\varepsilon_d=0$, $\Gamma_0=0.04$; (b) $U=\varepsilon_d=0$, $2V^2/(\pi D)=0.0006$. NRG parameters are the same as in Figs.\,3 and 4, respectively. The exact EOM results, calculated from Eqs.\,(24)-(26), are also presented for comparison.}
\end{figure}

Aside from being vulnerable to produce misleading results due to the involved $\delta\chi_c$ term, the NRG calculation of the impurity susceptibility $\chi_{\textrm{imp}}$ has another potential flaw in a more fundamental aspect. Note that the NRG logarithmic discretization is an approximate procedure transforming the continuous conduction bath into the discretized Wilson chain. The bath properties are qualitatively changed by this procedure due to the discretization error. With this respect, the NRG method is only suitable to calculate local quantities which do not explicitly involve the bath degrees of freedom, e.g., $\chi_\textrm{loc}$. The NRG calculation of any nonlocal quantity explicitly involving the bath degrees of freedom, e.g., $\delta\chi_c$ and thus $\chi_\textrm{imp}$, may be not reliable. A representative example is the spin susceptibility $\chi_c=(g\mu_B)^2\beta\langle S_b^2\rangle_{H_\textrm{bath}}$ of conduction electrons [i.e., the last term in Eq.\,(17)] in a flat wide band [$\rho(\varepsilon)=\rho_{_0}$], as shown in Fig.\,6. For the original continuous model, $\chi_c$ gives the temperature-independent Pauli paramagnetic susceptibility $\chi_c=\frac{1}{4}(g\mu_B)^2\rho_{_0}$. But the corresponding quantity in the discretized Wilson chain (calculated by the exact diagonalization method to highlight the discretization error) acquires a strong artificial temperature dependence (see Fig.\,6) due to the discretization error. This demonstrates that even for the flat wide band the NRG discretization error is important to nonlocal quantities. In the flat wide band case, since the nonlocal $\chi_\textrm{imp}$ is essentially a local quantity as $\chi_\textrm{imp}=\chi_\textrm{loc}$, the discretization error in $\chi_\textrm{imp}$ could be largely canceled by subtracting two nonlocal quantities (i.e., the total susceptibilities of the system with and without the impurity). This validates Wilson's definition in the flat wide band. However, such a cancellation of the discretization error may be not always strictly guaranteed in arbitrarily structured bands for which the impurity susceptibility $\chi_\textrm{imp}$ is a true nonlocal quantity, as shown in Fig.\,5. This again sheds a shadow on the NRG calculation of the impurity susceptibility.

\begin{figure}[]
\includegraphics[width=0.9\columnwidth]{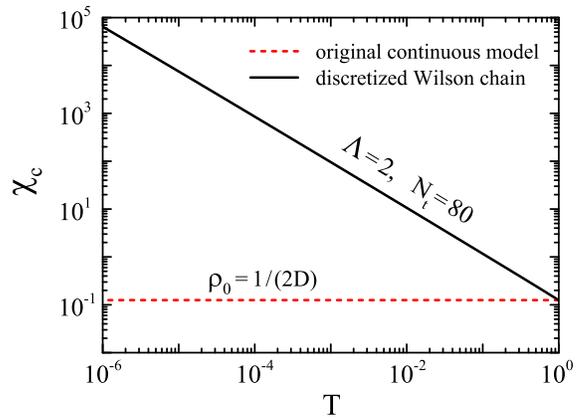}
\caption{(Color online) Spin susceptibility $\chi_c$ of conduction electrons in a flat wide band. For the original continuous model, $\chi_c$ is the temperature-independent Pauli susceptibility. For the discretized Wilson chain, $\chi_c$ is evaluated using the exact diagonalization method, and presented for $\chi_c/N_t$ with $N_t$ the total number of sites in the chain. We choose a sufficiently large number of sites to ensure that the coupling between the last two sites, $\sim\Lambda^{-(N_t-2)/2}$, is far less than the temperature.}
\end{figure}

\subsection{D. Remarks}
It is now in the position to critically discuss the effect of an energy-dependent conduction band on the magnetic response of an Anderson impurity. The results of Sec.\, III\,B and III\,C provide a definitive answer to the problem. As long as the host density of states is not gapped at the Fermi energy, for arbitrary energy dependence even though there are discontinuities or singularities in the band, the low-energy excitations of the impurity are always a Fermi liquid and paramagnetic. For such systems, the resultant Kondo screening of the impurity magnetic moment (demonstrated explicitly by the temperature dependence of $\chi_{\textrm{loc}}$) is qualitatively the same as in the flat wide band. The observation is clearly in contrast to the previous investigations \cite{Hofstetter1999,Mastrogiuseppe2014,Silva2006,Zhuravlev2011}. This is because the widely used impurity susceptibility $\chi_{\textrm{imp}}$, involving the additional susceptibility from the conduction electrons, is not a good quantity to determine the impurity Kondo physics when the host density of states is energy dependent. For example, a divergent and/or negative $\chi_{\textrm{imp}}$ as $T\rightarrow0$ certainly invalidates the standard definition of the Kondo temperature $T_0\equiv\lim_{T\rightarrow0}1/4\chi_{\textrm{imp}}$ \cite{Hewson1993,Bulla2008}. For such systems, the correct magnetic properties of the impurity must only be drawn from the local susceptibility.

Recently, Hanl and Weichselbaum \cite{Hanl2014} have proposed a new susceptibility $\chi^{\textrm{sc}}\equiv\chi_{\textrm{loc}}+\chi^{\textrm{mix}}$ for the Anderson impurity by adding the local susceptibility $\chi_{\textrm{loc}}$ and the so-called \emph{mixed} susceptibility
\begin{equation}
\chi^{\textrm{mix}}=2\lim_{B\rightarrow0}\frac{\partial}{\partial B}\langle M_\textrm{b}\rangle_{H+H'_m},
\end{equation}
in order to achieve universal Kondo scaling for narrow bandwidth. This mixed susceptibility, describing the bath magnetization in response to a local field, is a nonlocal quantity and thus may be still vulnerable to the discretization error. More importantly, by using the EOM approach, $\chi^{\textrm{mix}}$ can be expressed in the same form of $\delta\chi_c$, i.e., as the second term in the right-hand side of Eq.\,(20) only with the magnetic field now applied on the impurity not on the bath. This means $\chi^{\textrm{mix}}$ can still be negative for some structured bands. Therefore, there is no guarantee that in arbitrary energy-dependent bands the new susceptibility $\chi^{\textrm{sc}}$ is always reliable for characterizing the impurity Kondo effect, even though it works well for the narrow-band model.

\section{IV. Conclusion}
A comparative investigation of the local and impurity susceptibilities for an Anderson impurity, by using the FDM NRG technique, has demonstrated the importance of the local susceptibility in characterizing the intrinsic magnetic properties of the impurity. Within the FDM algorithm, while the calculation of the impurity susceptibility inevitably produces severe artifacts at very low temperatures, the local quantity, due to its sum-rule nature, can be calculated very accurately in all parameter regimes. In particular, the accuracy of the local susceptibility calculated at much lower computational cost is at least one order higher than that of the impurity susceptibility. For certain class of single-impurity Anderson systems in which the host density of states is arbitrarily energy-dependent but not gapped at the Fermi energy, we have revealed that the non-Fermi-liquid and/or diamagnetic behaviors found in the literature based on the knowledge of the impurity susceptibility are  spurious. The correct magnetic properties of the impurity in such systems should only be deduced from the local susceptibility, which suggests that the low-energy excitations of the impurity is always a Fermi liquid and paramagnetic. We hope this paper could indeed motivate more attention to the local susceptibility whenever the intrinsic magnetic response of the impurity is concerned.

\section{acknowledgments}
One of the authors, T.\,F. Fang, is grateful to T.\,A. Costi, A. Weichselbaum, and L. Merker for confirming the artifacts in the FDM algorithm to $\chi_{\textrm{imp}}$, H. Zhang for many inspiring discussions, and F.\,B. Anders for clarifying the phase factor in NRG matrix elements. This work is financially supported by NSF-China (11174115, 11325417, 11374362, and 11274364), 973 program of China (2012CB921704), NBRP of China (2012CB921303 and 2015CB921102), the fundamental research funds for central universities, and the research funds of Renmin University of China.

\begin{appendix}
\section{appendix: derivation of equation (12) for the local susceptibility}
\setcounter{equation}{0}
\renewcommand{\theequation}{A\arabic{equation}}
In this appendix, we provide the full details of deriving Eq.\,(12) within the FDM approach \cite{Weichselbaum2007}. We start from the spin correlation function $T(t)\equiv\langle[S_z(t),S^\dagger_z(0)]\rangle_H$ appearing in Eq.\,(11). Inserting the completeness relation (4) twice gives
\begin{eqnarray}
T(t)&=&\textrm{Tr}\left(S_z\rho e^{\frac{i}{\hbar }Ht}S_ze^{-\frac{i}{\hbar }Ht}-
\rho S_ze^{\frac{i}{\hbar }Ht}S_ze^{-\frac{i}{\hbar }Ht}\right)\nonumber\\
&=&\sum_{N,N^{\prime }=N_{\min }}^{N_{\max }}\sum_{l,l^{\prime },e,e^{\prime }}
\bigg [\left\langle l^{\prime
},e^{\prime };N^{\prime }\right\vert S_ze^{-\frac{i}{\hbar }Ht}\left\vert
l,e;N\right\rangle\nonumber \\
&&\qquad\quad\times\Big(\left\langle l,e;N\right\vert S_z\rho e^{\frac{i}{\hbar }Ht}\left\vert
l^{\prime },e^{\prime };N^{\prime }\right\rangle\nonumber\\
&&\qquad\quad-\left\langle l,e;N\right\vert \rho S_ze^{\frac{i}{\hbar }Ht}\left\vert
l^{\prime },e^{\prime };N^{\prime }\right\rangle\Big )\bigg ]\nonumber\\
&=&T^i(t)+T^{ii}(t)+T^{iii}(t),
\end{eqnarray}
where the double sum $\sum_{N,N'=N_{\textrm{min}}}^{N_{\textrm{max}}}$ is decomposed into three contributions with $N=N'$ ($T^i$ term), $N>N'$ ($T^{ii}$ term), and $N<N'$ ($T^{iii}$ term), respectively. The first contribution is
\begin{eqnarray}
T^i(t)&=&\sum_{N=N_{\min }}^{N_{\max }}\sum_{l,l^{\prime },e,e^{\prime }}
\bigg [\left\langle l^{\prime
},e^{\prime };N\right\vert S_ze^{-\frac{i}{\hbar }Ht}\left\vert
l,e;N\right\rangle\nonumber \\
&&\qquad\quad\times\Big(\left\langle l,e;N\right\vert S_z\rho e^{\frac{i}{\hbar }Ht}\left\vert
l^{\prime },e^{\prime };N\right\rangle\nonumber\\
&&\qquad\quad-\left\langle l,e;N\right\vert \rho S_ze^{\frac{i}{\hbar }Ht}\left\vert
l^{\prime },e^{\prime };N\right\rangle\Big )\bigg ].
\end{eqnarray}
The second contribution is
\begin{eqnarray}
T^{ii}(t)&=&\sum_{N^{\prime }=N_{\min }}^{N_{\max }-1}\sum_{N=N'+1}^{N_{\textrm{max}}}\sum_{l,l^{\prime },e,e^{\prime }}
\bigg [\nonumber\\
&&\quad\left\langle l^{\prime
},e^{\prime };N^{\prime }\right\vert S_ze^{-\frac{i}{\hbar }Ht}\left\vert
l,e;N\right\rangle\nonumber\\
&&\quad\times\Big(\left\langle l,e;N\right\vert S_z\rho e^{\frac{i}{\hbar }Ht}\left\vert
l^{\prime },e^{\prime };N^{\prime }\right\rangle\nonumber\\
&&\quad-\left\langle l,e;N\right\vert \rho S_ze^{\frac{i}{\hbar }Ht}\left\vert
l^{\prime },e^{\prime };N^{\prime }\right\rangle\Big )\bigg ]\nonumber\\
&=&\sum_{N=N_{\min }}^{N_{\max }-1}\sum_{k,l,e,e^{\prime }}
\bigg [\left\langle l,e^{\prime };N\right\vert S_ze^{-\frac{i}{\hbar }Ht}\left\vert
k,e;N\right\rangle\nonumber\\
&&\quad\quad\times\Big(\left\langle k,e;N\right\vert S_z\rho e^{\frac{i}{\hbar }Ht}\left\vert
l,e^{\prime };N\right\rangle\nonumber\\
&&\quad\quad-\left\langle k,e;N\right\vert \rho S_ze^{\frac{i}{\hbar }Ht}\left\vert
l,e^{\prime };N\right\rangle\Big )\bigg ].
\end{eqnarray}
The last equality of Eq.\,(A3) has applied the relation (6) and a notation change $N'\rightarrow N,\,l'\rightarrow l$ in the final result. The third contribution is
\begin{eqnarray}
T^{iii}(t)&=&\sum_{N=N_{\textrm{min}}}^{N_{\textrm{max}}-1}\sum_{N^{\prime }=N+1}^{N_{\max }}\sum_{l,l^{\prime },e,e^{\prime }}
\bigg [\nonumber\\
&&\quad\left\langle l^{\prime
},e^{\prime };N^{\prime }\right\vert S_ze^{-\frac{i}{\hbar }Ht}\left\vert
l,e;N\right\rangle\nonumber \\
&&\quad\times\Big(\left\langle l,e;N\right\vert S_z\rho e^{\frac{i}{\hbar }Ht}\left\vert
l^{\prime },e^{\prime };N^{\prime }\right\rangle\nonumber\\
&&\quad-\left\langle l,e;N\right\vert \rho S_ze^{\frac{i}{\hbar }Ht}\left\vert
l^{\prime },e^{\prime };N^{\prime }\right\rangle\Big )\bigg ]\nonumber\\
&=&\sum_{N=N_{\textrm{min}}}^{N_{\textrm{max}}-1}\sum_{k,l,e,e'}
\bigg [\left\langle k,e^{\prime };N\right\vert S_ze^{-\frac{i}{\hbar }Ht}\left\vert
l,e;N\right\rangle\nonumber \\
&&\quad\quad\times\Big(\left\langle l,e;N\right\vert S_z\rho e^{\frac{i}{\hbar }Ht}\left\vert
k,e^{\prime };N\right\rangle\nonumber\\
&&\quad\quad-\left\langle l,e;N\right\vert \rho S_ze^{\frac{i}{\hbar }Ht}\left\vert
k,e^{\prime };N\right\rangle\Big )\bigg ].
\end{eqnarray}
Again, the last equality of Eq.\,(A4) is due to the application of the relation (6). We substitute the full density matrix (7) into Eqs.\,(A2)-(A4), and then use the NRG approximation $e^{\pm\frac{i}{\hbar}Ht}\vert s,e;N\rangle\approx e^{\pm\frac{i}{\hbar}E^N_st}\vert s,e;N\rangle$, the orthonormality (5), and the local nature of the impurity spin $\langle s,e;N|S_z|s',e';N\rangle=\delta_{ee'}\langle s;N|S_z|s';N\rangle$. These lead to
\begin{eqnarray}
T^{i}\left( t\right)&=&\sum_{N=N_{\min }}^{N_{\max }}\sum_{l,l^{\prime }}\bigg \{e^{%
\frac{i}{\hbar }\left( E_{l^{\prime }}^{N}-E_{l}^{N}\right) t}\frac{%
4^{N_{\max }-N}}{Z}\nonumber\\
&&\times\left( e^{-\beta E_{l^{\prime }}^{N}}-e^{-\beta
E_{l}^{N}}\right)\left[ S_{z}\right] _{ll^{\prime }}^{N}\left[ S_{z}\right]
_{l^{\prime }l}^{N}\bigg\},
\end{eqnarray}
\begin{eqnarray}
T^{ii}\left( t\right)  &=&\sum_{N=N_{\min }}^{N_{\max }-1}\sum_{k,l}\bigg\{e^{\frac{%
i}{\hbar }\left( E_{l}^{N}-E_{k}^{N}\right) t}\frac{4^{N_{\max }-N}}{Z}\nonumber\\
&&\qquad\qquad\qquad\quad\times e^{-\beta E_{l}^{N}}\left[ S_{z}\right] _{kl}^{N}\left[ S_{z}\right]
_{lk}^{N}\bigg\}\nonumber \\
&&-\sum_{N=N_{\min }}^{N_{\max }-1}\sum_{N^{\prime }=N+1}^{N_{\max
}}\sum_{k,l,e}\bigg\{e^{\frac{i}{\hbar }\left( E_{l}^{N}-E_{k}^{N}\right) t}\left[
S_{z}\right] _{lk}^{N}\nonumber\\
&&\qquad\qquad\times\left\langle k,e;N\right\vert \rho _{N^{\prime
}}S_{z}\left\vert l,e;N\right\rangle\bigg\},
\end{eqnarray}
\begin{eqnarray}
T^{iii}\left( t\right) &=&\sum_{N=N_{\min }}^{N_{\max }-1}\sum_{N^{\prime
}=N+1}^{N_{\max }}\sum_{k,l,e}\bigg\{e^{\frac{i}{\hbar }\left(
E_{k}^{N}-E_{l}^{N}\right) t}\left[ S_{z}\right] _{kl}^{N}\nonumber\\
&&\qquad\qquad\qquad\times\left\langle
l,e;N\right\vert S_{z}\rho _{N^{\prime }}\left\vert k,e;N\right\rangle\bigg\}\nonumber\\
&&-\sum_{N=N_{\min }}^{N_{\max }-1}\sum_{k,l}\bigg\{e^{\frac{i}{\hbar }\left(
E_{k}^{N}-E_{l}^{N}\right) t}\frac{4^{N_{\max }-N}}{Z}\nonumber\\
&&\qquad\qquad\quad\,\times e^{-\beta E_{l}^{N}}%
\left[ S_{z}\right] _{kl}^{N}\left[ S_{z}\right] _{lk}^{N}\bigg\}.
\end{eqnarray}
Here the notation $[S_z]^N_{ss'}\equiv\langle s;N|S_z|s';N\rangle$ is introduced. The terms in Eqs.\,(A6) and (A7), which contain the $N'$th-shell density matrix $\rho_{N'}\equiv Z^{-1}\sum_{l,e}e^{-\beta E^{N'}_l}|l,e;N'\rangle\langle l,e;N'|$, need further calculations as follows:
\begin{eqnarray}
&&\qquad\qquad\qquad\sum_{e}\left\langle k,e;N\right\vert \rho _{N^{\prime }}S_{z}\left\vert
l,e;N\right\rangle\nonumber\\
&=&\sum_{N^{\prime \prime }=N_{\min }}^{N_{\max
}}\sum_{l^{\prime },e^{\prime },e}\left\langle k,e;N\right\vert \rho
_{N^{\prime }}\left\vert l^{\prime },e^{\prime };N^{\prime \prime
}\right\rangle\left\langle l^{\prime },e^{\prime };N^{\prime \prime
}\right\vert S_{z}\left\vert l,e;N\right\rangle\nonumber  \\
&=&\sum_{N^{\prime \prime }=N+1}^{N_{\max }}\sum_{l^{\prime },e^{\prime
},e}\left\langle k,e;N\right\vert \rho _{N^{\prime }}\left\vert l^{\prime
},e^{\prime };N^{\prime \prime }\right\rangle \left\langle l^{\prime
},e^{\prime };N^{\prime \prime }\right\vert S_{z}\left\vert
l,e;N\right\rangle\nonumber  \\
&=&\sum_{k^{\prime },e^{\prime },e}\left\langle k,e;N\right\vert \rho
_{N^{\prime }}\left\vert k^{\prime },e^{\prime };N\right\rangle \left\langle
k^{\prime },e^{\prime };N\right\vert S_{z}\left\vert l,e;N\right\rangle\nonumber  \\
&=&\sum_{k^{\prime }}\rho _{kk^{\prime }}^{NN^{\prime }}\left[ S_{z}\right]
_{k^{\prime }l}^{N}.
\end{eqnarray}
Similarly,
\begin{equation}
\sum_{e}\left\langle l,e;N\right\vert S_{z}\rho _{N^{\prime }}\left\vert
k,e;N\right\rangle=\sum_{k^{\prime }}\rho _{k^{\prime }k}^{NN^{\prime }}\left[ S_{z}\right]
_{lk^{\prime }}^{N}.
\end{equation}
In Eqs.\,(A8) and (A9), the reduced density matrix $\rho_{kk'}^{NN'}\equiv\sum_e\langle k,e;N|\rho_{N'}|k',e;N\rangle$. We can now collect all the contributions Eqs.\,(A5)-(A9) to obtain
\begin{eqnarray}
T\left( t\right)  &=&\sum_{N=N_{\min }}^{N_{\max }-1}\sum_{N^{\prime
}=N+1}^{N_{\max }}\sum_{k,k^{\prime},l}\bigg\{\Big[ e^{\frac{i}{\hbar }\left(
E_{k}^{N}-E_{l}^{N}\right) t}\nonumber\\
&&\qquad\qquad-e^{\frac{i}{\hbar }\left(
E_{l}^{N}-E_{k^{\prime }}^{N}\right) t}\Big]\rho _{k^{\prime
}k}^{NN^{\prime }}\left[ S_{z}\right] _{kl}^{N}\left[ S_{z}\right]
_{lk^{\prime }}^{N}\bigg\}\nonumber \\
&&+\sum_{N=N_{\min }}^{N_{\max }-1}\sum_{k,l}\bigg\{\left[ e^{\frac{i}{\hbar }%
\left( E_{l}^{N}-E_{k}^{N}\right) t}-e^{\frac{i}{\hbar }\left(
E_{k}^{N}-E_{l}^{N}\right) t}\right]\nonumber\\
&&\qquad\qquad\qquad\times\frac{4^{N_{\max }-N}}{Z}e^{-\beta
E_{l}^{N}}\left[ S_{z}\right] _{kl}^{N}\left[ S_{z}\right] _{lk}^{N}\bigg\}\nonumber \\
&&+\sum_{N=N_{\min }}^{N_{\max }}\sum_{l,l^{\prime }}\bigg\{e^{\frac{i}{\hbar }%
\left( E_{l^{\prime }}^{N}-E_{l}^{N}\right) t}\frac{4^{N_{\max }-N}}{Z}\nonumber\\
&&\quad\times\left( e^{-\beta E_{l^{\prime }}^{N}}-e^{-\beta E_{l}^{N}}\right) \left[
S_{z}\right] _{ll^{\prime }}^{N}\left[ S_{z}\right] _{l^{\prime }l}^{N}\bigg\}.
\end{eqnarray}

Substituting Eq.\,(A10) into Eq.\,(11) and performing the Fourier transformation by using the following integral representation of the Heaviside step function
\begin{equation}
\Theta(t)=\frac{1}{2\pi i}\int_{-\infty}^\infty\frac{e^{it\tau}}{\tau-i\eta}\textrm{d}\tau
\end{equation}
with $\eta\rightarrow0^+$, we obtain the retarded Green's function of impurity spin
\begin{equation}
G_{S_z}(\varepsilon)=G_{S_z}^i(\varepsilon)+G_{S_z}^{ii}(\varepsilon)+G_{S_z}^{iii}(\varepsilon),
\end{equation}
\begin{eqnarray}
G_{S_{z}}^{i}\left( \varepsilon \right) &=&\sum_{N=N_{\min }}^{N_{\max
}-1}\sum_{N^{\prime }=N+1}^{N_{\max }}\sum_{k,k^{\prime },l}\Bigg\{\bigg[ \frac{1}{%
\varepsilon -\left( E_{l}^{N}-E_{k}^{N}\right) +i\eta }\nonumber\\
&&-\frac{1}{\varepsilon
-\left( E_{k}^{N}-E_{l}^{N}\right) +i\eta }\bigg] \rho _{k^{\prime
}k}^{NN^{\prime }}\left[ S_{z}\right] _{kl}^{N}\left[ S_{z}\right]
_{lk^{\prime }}^{N}\Bigg\},\nonumber\\
\end{eqnarray}
\begin{eqnarray}
G_{S_{z}}^{ii}\left( \varepsilon \right)&=&\sum_{N=N_{\min }}^{N_{\max
}-1}\sum_{k,l}\Bigg\{\Bigg[ \frac{4^{N_{\max }-N}Z^{-1}e^{-\beta E_{l}^{N}}}{
\varepsilon -\left( E_{k}^{N}-E_{l}^{N}\right) +i\eta }\nonumber\\
&&-\frac{4^{N_{\max
}-N}Z^{-1}e^{-\beta E_{l}^{N}}}{\varepsilon -\left(
E_{l}^{N}-E_{k}^{N}\right) +i\eta }\Bigg] \left[ S_{z}\right] _{kl}^{N}%
\left[ S_{z}\right] _{lk}^{N}\Bigg\},\nonumber\\
\end{eqnarray}
\begin{eqnarray}
G_{S_{z}}^{iii}\left( \varepsilon \right)&=&\sum_{N=N_{\min }}^{N_{\max
}}\sum_{l,l^{\prime }}\Bigg\{\frac{4^{N_{\max }-N}}{Z}\frac{e^{-\beta E_{l^{\prime
}}^{N}}-e^{-\beta E_{l}^{N}}}{\varepsilon -\left( E_{l}^{N}-E_{l^{\prime
}}^{N}\right) +i\eta }\nonumber\\
&&\qquad\qquad\times\left[ S_{z}\right] _{ll^{\prime }}^{N}\left[ S_{z}%
\right] _{l^{\prime }l}^{N}\Bigg\}.
\end{eqnarray}
In deriving Eq.\,(A13), the Hermitian conditions $\rho^{NN'}_{kk'}=\rho^{NN'}_{k'k}$ and $[S_z]^{N}_{lk}=[S_z]^{N}_{kl}$ are used. Substituting the above Lehmann representation for $G_{S_z}(\varepsilon)$ into Eq.\,(10) yields straightforwardly Eq.\,(12) in the main text.

Finally, we would like to remind the reader that the $N'$th-shell density matrix $\rho_{N'}$ and hence the reduced density matrix $\rho^{NN'}_{kk'}$ defined here are different by a factor $Z^{-1}4^{N_{\textrm{max}}-N'}\sum_le^{-\beta E^{N'}_l}$, as compared with the original definitions in Ref.\,[17]. Nevertheless, the numerical algorithm for $\rho^{NN'}_{kk'}$ remains the same as before.
\end{appendix}

\end{document}